\documentstyle[aps,prl]{revtex}
\twocolumn
\draft
\begin{document}
\title{Nucleation and Growth of Normal Phase in Thin Superconducting Strips}
\author{I. Aranson $^1$, B.Ya. Shapiro $^1$ and V. Vinokur $^2$}
\address{
$^1$ Department of Physics and Jack and Pearl Resnick Institute of Advanced
Technology, \\
Bar Ilan University, Ramat Gan 52900, Israel\\
$^2$ Argonne National Laboratory,
9700 South  Cass Avenue, Argonne, IL 60439
}
\date{\today}

\maketitle
\begin{abstract}

We investigate the kinetics of  normal phase nucleation and flux line
condensation in the type-II superconductors by numerical study of the
time-dependent Ginzburg-Landau equation.
We have shown that under the sufficient transport current the normal phase
nucleates in the superconducting strips in a form of the macroscopic droplets
having the multiple topological charge. We discuss the stability
and the dynamics of the droplets. We found that pinning suppresses the droplet
formation.

\end{abstract}

The study of magnetic flux penetration in type II superconductors has
attracted a wide interest both in view  of important technological questions
and as an prototype of a general class of problems of nonlinear dynamics.
Observations showed that the flux dynamics exhibits features that are similar
to the viscous-fingering growth phenomenon in liquid-solid systems
\cite{dors,Huebener,vlasko,duran,inden}.  In particular recent experiments
revealed dendrite flux penetration and the fingering of the remagnetization
front. \cite{vlasko,duran,inden}.
The formation of vortex structure is traditionally viewed as the sequential
penetration of vortices through the Bean-Livingstone surface barrier
\cite{burl}.
It was found recently that flux penetration may also occur  via dynamic
instabilities of order parameter caused by the applied current and/or magnetic
field.  Numerical simulations revealed the invasion of the extended
macroscopic normal areas (droplets) carrying flux into the superconducting
sample \cite{ags,kaper}.

While the formation of normal areas looks natural for type I superconductors
with the positive surface energy of the normal-superconductor (NS) interface,
it seems surprising at the first sight that such an interface having in the
static case the {\it negative} surface energy persists in type II
superconductors.  We see the explanation of this phenomenon in the
fact that the transport current or alternating magnetic field,
drives the superconductor into a strongly non-equilibrium state where the {\it
moving} interface becomes stable.  The idea that the free energy
considerations do not apply to  nonstationary processes in superconductors was
put forward by Anderson et al \cite{Abrik} in the context of phase-slips
phenomenon.

In this letter we report on our investigation of the kinetics of {\it normal}
phase nucleation  and flux lines condensation
in the type-II superconductors. We present the results of a numerical study of
the dynamics of the flux penetration into strips with the transverse
dimensions less than the effective penetration length
$\lambda_{eff}={\lambda}^2/h$, where $h$ is the thickness of the strip and
$\lambda$ is the London penetration depth.
We propose that the existence of the macroscopic normal regions is the direct
consequence of their motion under the transport current. A current  cannot
penetrate the {\it immobile} compact normal zone immersed into a
superconductor \cite {Abrik}, therefore the NS interface moves towards normal
phase with the velocity going to infinity \cite{kulik,ivlev}, and the normal
droplet disappears. At the same time, the penetration of the current into a
normal phase makes it stable with respect to small fluctuations \cite{gorkov},
i.e. the transport current drives the system into a {\it bistable} state.
Since the expulsion of the current from the normal regions requires a finite
time, the current penetrates the moving normal droplet.  The normal state
develops and invades into the superconducting
region provided the current $j$ flowing through the interface exceeds the
stall current $j^{*}$ \cite
{Kramer,lichar}.
Thus the sufficient transport current stabilizes moving nuclei of the normal
state in type II superconductors.

The process of flux penetration occurs via the depression of the order
parameter on the macroscopic scale and can be viewed as the nucleation of the
extended droplets of the normal phase in the superconducting sample.  An
adequate description of such process involving the fast variations of the
order parameter on the relevant spatial scale is given by the time-dependent
Ginzburg-Landau equation (TDGLE) completed by the appropriate Maxwell
equations:
\begin{eqnarray}
u (\partial_t+i\mu)\Psi &=& \left(\nabla-i{\bf A}\right)^2\Psi +
\left(1-\left|\Psi\right|^2\right)\Psi ,  \label{eq1_1} \\
{\bf j}&=& \left|\Psi\right|^2\left(\nabla\varphi-{\bf A}\right) -
\left(\nabla\mu+\partial_t{\bf A}\right),  \label{eq1_2} \\
\nabla\cdot{\bf j} &=& 0 \;\;\;\;,\; \nabla\cdot{\bf A} = 0\;,  \label{eq1_5}
\\
\Delta {\bf A} &=& - \frac{1}{\lambda_{eff}} {\bf j} \delta(z)
\label{eq1_6}
\end{eqnarray}
where $\Psi$ is the (complex) order parameter, $\varphi = arg\Psi$, ${\bf A}$
and $\mu$ are vector and scalar potentials, and ${\bf j}$ is the current
density. The value of the dimensionless material parameter  $u$ is obtained
from the microscopic
theory \cite{gorkov}. The unit of length is the coherence length $\xi$, unit
of time is $t_0={\xi}^2/Du$, $D=v_Fl/3$ is the diffusion constant, $l$ is a
mean free path, $v_F$ is a Fermi velocity, the field is measured in  units of
the upper critical field $H_{c2}=\Phi_0/2\pi\xi^2$, $\Phi_0$ is the flux
quantum. The unit of current is $j_0=\sigma\hbar/2et_0$, where $\sigma$ is the
normal conductivity.  In these units the depairing current
$j_p=2/3\sqrt{3}\approx 0.3875$.
The condition $\lambda_{eff} \gg 1$ enables us to neglect the magnetic field
created by
currents \cite{pearl} and therefore drop the equation (\ref{eq1_6}). We
choose the origin of the coordinate frame at the mid-point of the strip with
the $x$ axis lengthwise and the $y$ axis in the lateral direction, so that
the edges are located at $(x,-d/2)$ and $(x,d/2)$. The normal to the strip
magnetic field $B$ is associated with the vector potential ${\bf A}=(By,0,0)$.

We performed  numerical simulations of TDGLE. We took a homogeneous
superconducting state initial conditions ($\Psi=1$, i.e. the state without
magnetic field)
perturbed by a small amplitude noise. We used the no-flux boundary conditions,
$\partial_y \Psi =0$ (i.e. the boundary with vacuum) in
the transverse direction and the NS boundary conditions in the longitudinal
direction ($\Psi(x,y)
\to 0$ for $x \to 0, L$, where $L$ is the strip length).
We apply the split-step method described in
\cite{ags,AKW}, the number of the grid points was 256$\times$256 and the
time-step was $0.05-0.1$. Results of the simulations are shown in Fig. 1. The
simulations were performed for $j=0.25$, $B=0.0175$ where as it had been shown
in \cite{ags}, the pure superconducting state is unstable with respect to
vortex nucleation (note that our equations do not contain fluctuations). The
integration domain
was $120\times 60$.

On the Fig. 1 the large dark droplets (for $t=50,80$)
represent the normal phase emerging at the one side of the strip
and traversing toward the opposite edge. The droplets are the long-living
objects and as well as vortices
play a crucial role in the dissipative processes.
In our simulations the topological charge of these droplets would become as
big as  5 -7 and
even more. The droplets possess long tails (due to a finite relaxation time of
the order parameter at the superconducting areas swept by the droplet).
Our simulations show
that new vortices appear at the edge just at  the tail and then get sucked
in the droplet.
This can be easily understood since the formation of the new vortices is
favored in the regions with
suppressed order parameter.
The normal phase areas can evolve in two different ways.
First, the normal droplet emerges at the edge, passes through the sample and
vanishes  at the opposite edge of the strip.
In the second scenario that occurs under elevated currents the droplet
traverses a strip leaving a channel (wake) of the normal phase behind.

This scenario  is shown on
Fig. 1, $t=260$.
Than this channel traversing the sample breaks into the sequence of
vortices (vortex street), which then propagate across the strip and annihilate
at the edge.
The nucleation and the  propagation of the droplets
and the vortices gives rise to non-periodic voltage oscillations
along the strip.

The droplets posses the topological charge $n$ proportional to the gain in the
superconducting phase along the loop enclosing the normal area. A relationship
between the characteristic size $R$ of the  nucleus and $n$ is determined then
from the condition that the supercurrent encircling the nucleus ($\sim n/R$)
becomes equal to the $j_p$ giving $R \sim n/j_p$.
The size of the droplets
in the strip can be estimated from the condition that total transport current
at the distance $R$ from the edge
$j(R)  \approx j - B (R-d/2) $ is equal to the $j_p$.
It gives $R = d/2+(j-j_p)/B$.
For the chosen parameters we obtain $R \approx 20$ and $n \approx 6 \div 7$,
which is in qualitative agreement with the results of simulations.
We expect that the above consideration holds also for the large, $d\gg\lambda$
samples where $\lambda$ takes the role of the characteristic length. We
observed that droplets move much faster than single vortices. Simple analysis
shows that the Magnus force exerted on the droplet grows
linearly with $n$ whereas the mobility saturates for large $n$, resulting in
the velocity growth.


Shown on Fig. 1 is a sequence of snapshots demonstrating a
remagnetization process (we reversed the direction of the magnetic field at
$t=200$).
At the first stage of remagnetization large normal phase areas develop at the
edge of the
strip. These areas swallow vortices
corresponding to the previous direction of the magnetic field. Then the normal
areas assume more complicated form and then break up into smaller droplets.
For the
zero applied current
the Abrikosov  vortex lattice is formed
in the external field. In contrast to
the case with nonzero applied current, vortices penetrate from both edges of
the strip.
When the direction of the field is reversed,
large normal areas develop
at both edges and swallow vortices corresponding to the initial direction of
the field.
After a while the new Abrikosov lattice forms with vortices along the
reversed direction of the field.

To include the Hall effect in our simulations
we introduce the complex material parameter $u=5.79+i$. The
imaginary correction to $u$ describes the effect of the transverse Hall force
on the vortices drift \cite{dorsey}.  This gives rise to the Hall voltage,
moreover, we observe the turn of the droplet tail. We suggest that the
rotation
of the droplet's tail in the experimental work \cite{vlasko} is caused
by a significant Hall contribution.

To summarize, we have found the long living droplets of normal phase inside a
superconducting phase, and observed that they may posses the topological
charge than can significantly exceed unity. Note that the droplets must be
distinguished from the
 Abrikosov vortices with multiple charge. The linear
stability analysis shows that such vortices are unstable with respect to
the splitting into single charged vortices. The characteristic time of the
splitting is of
about $10-15$ dimensionless units and, therefore, cannot explain the existence
of long-living droplets. Note that these droplets may be viewed as the result
of the "fusion" of the separate vortices.

The qualitative arguments describing the droplet dynamics can be put on the
more rigorous basis for the droplets with the size well exceeding
the coherence length $\xi$. In this case the boundary of the droplet can
be considered locally as slightly curved NS interface. Inside the droplet
the order parameter $\Psi $ vanishes and the field is described entirely by
Laplace equation
\begin{equation}
\Delta \mu =0
\label{mu}
\end{equation}
The Eq. \ref{mu} has to be completed by the boundary conditions on the
interface, deduced from the continuity equation $\nabla {\bf j}=0$. It gives
the relation between the components of
currents normal to the interface $j_n^{(n)}=j_n^{(s)}$, where
superscripts $s,n$ denotes currents in normal and superconducting regions
respectively. Using Eq. (\ref{eq1_2}) we arrive at the first boundary
condition $
-\nabla _n\mu ^{(n)}=|\Psi |^2(\nabla _n\varphi -A_n)-\nabla _n\mu ^{(s)}
$
(here $\nabla _n$ means normal projection of the gradient).
The order parameter in superconducting region near the slightly curved
interface
is given in "adiabatic approximation" by $|\Psi|^2 = 1-(\nabla \varphi
_0-A)^2$.

The phase $\varphi$ of the superconducting order parameter in the leading
order
is described by Laplace equation
\begin{equation}
\Delta \varphi=0  \label{lap}
\end{equation}
together with the equation
for normal velocity of the interface. They latter can be derived from Eq.
(\ref{eq1_1}) for the slightly curved interface. The small curvature $\chi $
renormalizes the
normal velocity $c_n$  of the interface according Gibbs-Thomson condition
$
c_n=c_0-\chi  $,
where $c_0$ is the velocity of flat interface.

For the flat NS interface the velocity $c_0(j)$ is a
function of the transport current.  The one-dimensional situation had been
considered in \cite{lichar}, where the existence of the "stall" current $j^*$
at which  the interface velocity becomes equal to zero had been established.
For $u=5.79$ the stall current was found to be $j^{*}=0.335$, and $ c_0(j\to
j^*)\approx\alpha (j^*-j)$, where $\alpha=0.6$ is the numerical factor.
In two dimensional situation the topological charge of the droplet induces the
circular current $%
j_\tau$ tangential to the interface which modifies its the velocity.
To account for  the effect of the tangential current we take the order
parameter close to the nearly flat interface in a form (the interface is
parallel to $y$-axis, and we use a frame moving together with the interface
along $x$-axis with the velocity $c$)
$\Psi = F(x-ct) \exp[ i k_y y +\phi(x-ct)] $, where $k_x = \lim_{x\to -\infty}
\phi_x $.
and $j_\tau =(1-k_y^2-k_x^2)k_y$, $j_n=(1-k_x^2-k_y^2)k_x$.
A simple scaling analysis shows that the
current renormalizes the interface velocity  $c$ as
\begin{equation}
c(j_n,j_\tau )= c_0(\tilde j) \sqrt{1-k_y^2},  \label{vel}
\end{equation}
where $\tilde j =j_n/(\sqrt{1-k_y^2})^3$.
If the curvature of the interface is small (i.e $\chi \simeq 1/R\ll 1$), the
interface itself is defined by the additional
condition that at the (flat) interface $\mu =\mu
_0=k_x c(j_n,j_\tau )$. After that the problem is completely defined.

In the superconducting phase we have the  Eq.  (\ref{lap}) completed by the
boundary conditions for $\varphi $ on the strip edges.
Thus the problem under
study is a generalization of well-know problem of the Laplacian growth (see,
e.g. \cite{kessler}).
A new feature is that the function $%
\varphi $ is a multivalued one and has branch cuts. This multivalueness means
that the obtained equations contain implicitly vortex solutions: vortices can
appear and/or vanish via the formation of singularity at the interface.  The
detailed consideration of these equations we leave for the future, for now we
would like to mention that the
linear
stability analysis shows that the flat interface with the current flowing
through is stable with respect to small perturbations.  The above discussion
and the results of our simulations make us to conclude that the passage of the
current suppresses the NS interfaces
instabilities in
thin superconducting films.

To study the effects of pinning we carried out
simulations of TDGLE with randomly distributed pinning centers.  In the
presence of the weak pinning the newly formed droplets assume the "fractal"
configuration
since the normal phase tries to settle at the pinning sites where the order
parameter is already suppressed (see Fig. 2). The moving droplets percolates
along the easy paths connecting the pinning sites, but
the pinning centers impede the interface motion.  As a result the current that
penetrates the normal area gets smaller and cannot support the existence of
the droplet any more and droplets break up.
For stronger pinning the droplets do not form at all, and single vortices
penetrate the strip via the jumping resembling the vortex motion through the
array of linear defects \cite{nelson}.

Finally we discuss briefly the time scale of the observed effects.  The
characteristic time in dirty superconductors is
$t_0\simeq\hbar/T_c(1-T/T_c)\approx 10^{-14}\div10^{-11}$sec depending on the
temperature interval.  This means that the considered phenomena develop on the
nanosecond scale. However the process of the flux penetration can be
considerably slowed down by pinning. In this case the characteristic time (for
"dendrite" formations for example) should include macroscopic characteristics
such as the size of the sample and the average pinning strength \cite{vinokur}
and can grow up considerably.

In conclusion, we have shown that under the sufficient transport current the
normal phase nucleates in the superconducting strips in a form of the
macroscopic droplets which tear off at the edges and further propagate across
the sample.  These droplets possess the multiple topological charge related to
the magnetic flux they carry.  Pinning suppresses the droplet formation
converting normal area into the multi-connected fractal formations which then
split into the separate vortices.
We believe that the observed phenomena are not specific to the thin strips,
and that the same mechanism governs the normal phase formation in large
samples as well.

We are grateful to A. Koshelev and U. Welp for illuminating discussions.
This work was supported through U.S. Department of Energy, BES-Materials
Sciences, under contract \# W-31-109-ENG-38.  The work of IA and BS was partly
supported by the Raschi Foundation and ISF. IA acknowledges the support by NSF
office of the Science and Technology Center under contract \# DMR 91-20000 at
Argonne National Laboratory.  The visit of VV to Israel was supported by the
Rich Foundation via the Israel Ministry of Science and Arts.

\references{
\bibitem{dors}H. Frahm, S. Ullah,  and A. Dorsey, Phys. Rev. Lett.,
{\bf 66}, 3067 (1991).
F. Liu, M. Mondello, and N. Goldenfeld, Phys. Rev. Lett.,
{\bf 66}, 3071 (1991). R.E. Goldstein, D. Jackson, and A.T. Dorsey,
cond-mat/9411007, 1994.

\bibitem{Huebener} R.P. Huebener, {\it Magnetic Flux Structures in
Superconductors},
(Springer-Verlag, New York, 1979).
 \bibitem{vlasko}  V.V. Vlasko-Vlasov et al,
Physica C, {\bf 222}, 361 (1994). \bibitem{duran} C.A. Duran,
P.L. Gammel, R.E. Miller, and D.J. Bishop, \prb, {\bf 52}, 75 (1995).
\bibitem{inden} M.V. Indenbom et al,
Phys. Rev. B, {\bf 51}, 15484 (1995).
\bibitem{burl} L. Burlachkov et al, \prb, {\bf 50}, 16770 (1994).
\bibitem{ags} I. Aranson, M. Gitterman, and B. Ya. Shapiro, \prb, {\bf 51},
3092, (1995).
\bibitem{kaper} H. Kaper et al, to be publised.

\bibitem{Abrik}
A.A. Abrikosov, {\it Fundamentals of the Theory of Metals},
(Elsevier, New York,  1988).

 \bibitem{kulik} I.O. Kulik, Zh. Eks. Teor. Fiz., {\bf 59}, 584, (1970);
[Sov. Phys.-JETP, {\bf 32} , 318, (1971)].
\bibitem{ivlev} B.I. Ivlev and N.B. Kopnin, Adv. Phys, {\bf 33}, 47 (1984).

\bibitem{gorkov} L.P. Gor'kov
and N.B. Kopnin, Usp. Fiz. Nauk, {\bf 116}, 413 (1975).
\bibitem{Kramer} R.J. Watts-Tobin, Y. Kr\"ahenb\"uhl,  and L. Kramer,
J. of Low. Temp. Phys., {\bf 42}, 459 (1981).
\bibitem{lichar} K.K. Likharev, Zh. Eks. Teor. Fiz. Pis. Red. {\bf 20}, 730,
(1974);
[Sov. Phys.-JETP Letters, {\bf 20}, 338 (1974)].

\bibitem{pearl} J. Pearl, Appl. Phys. Lett, {\bf 5}, 65 (1966).

\bibitem{AKW} I. Aranson, L. Kramer, and A. Weber, J. Low. Temp. Phys, {\bf
89},859 (1992).
\bibitem{dorsey}
A. T. Dorsey, \prb, {\bf 46}, 8376 (1992);
A.G. Aronov et. al, \prb, {\bf 51}, 3880 (1995).

Phys., {\bf 97}, 215 (1994).
\bibitem{kessler} D. Kessler, J. Koplik, and H. Levine, Adv. Phys., {\bf 37},
225 (1988).

\bibitem{nelson}D. R. Nelson and V. M. Vinokur, \prl, {\bf 68}, 2398 (1992).
\bibitem{vinokur}V.M.Vinokur, M.V.Feigelman, and V.B.Geshkenbein, Phys. Rev.
Lett., {\bf 67}, 915 (1991)
}
\begin{figure}
\caption{Dynamics of the normal phase.
The current is applied along
$x$-axis and the magnetic field is perpendicular to the strip.
Gray-coded images show $|\Psi(x,y)|$ (
$|\Psi|=0$ is shown in black and $|\Psi|=1$ is shown in white).
The field is reversed  at  $t=200$.
}
\end{figure}

\begin{figure}
\caption{Normal phase penetration at $t=40$,
$j=0.25$, $B=0.018$ in the presence of 180 randomly distributed pinning
centers,
other parameters are the same as in Fig. 1.
}
\end{figure}

\end{document}